\documentclass{article}

\usepackage{arxiv}

\usepackage[utf8]{inputenc}
\usepackage[T1]{fontenc}
\usepackage{lmodern}

\usepackage{hyperref}
\hypersetup{hypertexnames=false}
\usepackage{xurl}
\usepackage{booktabs}
\usepackage{amsfonts}
\usepackage{amsmath}
\usepackage{amssymb}
\usepackage{mathtools}
\usepackage{graphicx}
\usepackage{microtype}

\usepackage{algorithm}
\usepackage{algpseudocode}

\usepackage{enumitem}
\usepackage{listings}
\usepackage{multirow}

\usepackage[numbers,compress]{natbib}
\usepackage{doi}
\usepackage{placeins}

\usepackage{xcolor}
\usepackage{tikz}
\usetikzlibrary{positioning,calc,arrows.meta,shadows}
\usepackage{fontawesome5}

\definecolor{layerBase}{HTML}{EBF3FB}
\definecolor{layerKernel}{HTML}{DCE9F7}
\definecolor{layerArtifact}{HTML}{FFF4D6}
\definecolor{layerGov}{HTML}{F9E4D8}
\definecolor{textDark}{HTML}{202124}

\tikzset{
	layer_box/.style={
		rectangle,
		rounded corners=4pt,
		draw=textDark!32,
		line width=0.7pt,
		text centered,
		align=center,
		text=textDark,
		font=\small\sffamily,
		drop shadow={opacity=0.06, shadow xshift=0.045cm, shadow yshift=-0.045cm},
		inner sep=5.5pt
	},
	core_box/.style={
		rectangle,
		rounded corners=4pt,
		draw=textDark!40,
		line width=0.95pt,
		text centered,
		align=center,
		text=textDark,
		font=\small\sffamily\bfseries,
		drop shadow={opacity=0.08, shadow xshift=0.05cm, shadow yshift=-0.05cm},
		inner sep=6pt
	},
	gov_arrow/.style={-{Stealth[scale=1.0]}, draw=textDark!68, line width=1pt},
	trace_arrow/.style={-{Stealth[scale=0.95]}, draw=textDark!58, line width=0.82pt, dashed}
}

\title{Governed Evolution of Agent Runtimes through Executable Operational Cognition}

\newif\ifuniqueAffiliation
\uniqueAffiliationtrue

\ifuniqueAffiliation % Standard variant of author block
\author{ \href{https://orcid.org/0009-0008-0201-2984}{\includegraphics[scale=0.06]{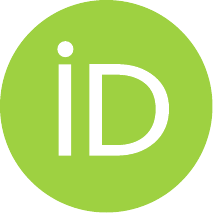}\hspace{1mm}Mariano Garralda-Barrio}\thanks{Independent Researcher / Investigador Independiente.} \\
Independent Researcher\\
Lleida, Spain \\
\texttt{mariano.garralda.r@gmail.com} \\
}
\fi

\renewcommand{\headeright}{}
\renewcommand{\undertitle}{A Systems Vision Paper}

\begin{document}

	\maketitle

	\begin{abstract}
		Recent advances in agentic systems increasingly treat code as an executable operational substrate rather than as a disposable output artifact. Prior work such as \emph{Code as Agent Harness} frames validated agent-generated artifacts as runtime entities that can be created, executed, revised, persisted, and reused within long-running cognitive loops. However, the governance, lifecycle management, and operational evolution of such artifacts remain under-specified.

		This paper proposes a framework for governed runtime evolution in multi-agent systems through executable operational cognition. We formalize agent-generated artifacts as persistent runtime capabilities that progressively become part of the operational substrate rather than transient intermediate outputs. Building on this perspective, we introduce \emph{HarnessMutation} as a governed mechanism for lifecycle-aware runtime adaptation operating under explicit validation, traceability, evaluation, and rollback constraints.

		Rather than treating runtime adaptation as unrestricted self-modification, the proposed framework models evolution as a bounded and observable process over persistent operational memory. It further shows how these ideas can be operationalized over modern agent runtimes and governance-oriented orchestration systems, providing a conceptual foundation for adaptive infrastructures whose evolution remains explicit, auditable, and constrained.
	\end{abstract}

	\section{Introduction}

	Large language models have accelerated the development of agentic systems capable of planning, retrieval, tool execution, code generation, verification, and iterative refinement, including program-aided reasoning, reasoning--acting agents, code-grounded evaluation, and software-engineering agents \citep{chen2022program,yao2023react,gao2023pal,jimenez2024swebench,openhands}. Early agent systems were often described primarily in terms of model reasoning and tool use. More recent systems, however, expose a deeper pattern: the operational substrate surrounding the model is becoming central to agent reliability.

	This shift creates a governance problem. As agentic runtimes accumulate generated workflows, prompts, evaluators, routing policies, executable skills, and other validated artifacts, those artifacts may begin to influence future behavior. Prior work increasingly supports such capability accumulation through lifelong skill libraries, prompt and context evolution, and harness optimization \citep{wang2023voyager,agrawal2025gepa,zhang2025agenticcontext,lou2026autoharness,lee2026metaharness,yang2026skillopt}, but it does not directly specify how accumulated artifacts should remain stable, auditable, reversible, and operationally safe over time. Without governance-aware evolution mechanisms, runtimes risk capability drift, silent regressions, evaluation contamination, and progressively non-auditable operational behavior.

	This work shifts the focus from code as an execution medium toward code as a governed evolutionary substrate for persistent operational adaptation under explicit runtime constraints. The proposed framework naturally extends to multi-agent systems where distinct agents specialize in generation, validation, evaluation, governance, and execution of operational artifacts. Under this interpretation, runtime evolution becomes a coordinated distributed process rather than a monolithic self-evolving loop.

	The survey \emph{Code as Agent Harness} frames this shift explicitly. It argues that code is no longer only a target output of language models, but an executable, inspectable, and stateful medium through which agents reason, act, observe feedback, and verify progress \citep{ning2026codeagentharness}. In that view, long-running agentic systems involve three coupled elements: model-internal capabilities, system-provided harness infrastructure, and agent-initiated code artifacts. The last category is particularly important for this paper. Agent-initiated code artifacts are interactive code objects that agents create, execute, observe, revise, persist, and share within the task execution loop. Examples include regression tests, temporary tools, domain-specific programs, executable workflows, reusable skills, and intermediate program states.

	This paper takes that distinction as its starting point and asks a further systems question: if agent-initiated executable artifacts increasingly participate in reasoning, action, verification, memory, and coordination, when and how do such artifacts become reusable runtime capabilities, and how should their evolution be modeled, governed, and operationally constrained? Existing work has identified the centrality of code artifacts inside the harness loop. The contribution of this paper is to formalize these artifacts as an optimization, lifecycle, and governance substrate.

	We propose the notion of \emph{executable operational cognition}. Under this view, generated code artifacts are not merely temporary outputs produced during a task. When persisted, evaluated, versioned, composed, and reused, they become operational capabilities that shape future runtime behavior. This reframes agent memory from passive retrieval toward executable capability accumulation.

	The resulting architecture is harness-oriented. Rather than modeling agents as loosely specified autonomous entities, we model executable harnesses as governed operational configurations composed of prompts, tools, evaluators, memory, policies, workflows, and persistent code artifacts. Iterative self-improvement then becomes an optimization process over harness configurations and executable cognition artifacts. Accordingly, this paper is a conceptual and architectural systems paper: its goal is to define a coherent design space and governance model for future empirical implementations.

	\subsection{Terminology and Conceptual Levels}

	To avoid ambiguity, the framework distinguishes between three conceptual levels participating in runtime evolution.

	An \emph{artifact} denotes an individual generated operational entity such as a prompt, evaluator, workflow, routing policy, executable skill, or code component produced during runtime execution.

	A \emph{capability} denotes an artifact that has successfully passed validation, governance, and persistence stages, becoming a reusable operational component integrated into future runtime behavior.

	Finally, \emph{operational cognition} refers to the emergent system-level behavior produced through the coordinated interaction, execution, mutation, governance, and composition of persistent capabilities across the operational substrate.

	\subsection{Contributions}

	The paper makes the following conceptual and architectural contributions:

	\begin{itemize}
		\item \emph{Executable operational cognition}: a framing in which agent-generated executable artifacts become persistent runtime capabilities.

		\item \emph{HarnessMutation}: a governed transformation mechanism for adapting prompts, workflows, evaluators, routing policies, skills, and runtime behaviors.

		\item \emph{Lifecycle-governed runtime evolution}: a model for validating, promoting, deprecating, and reusing operational artifacts under explicit governance constraints.

		\item \emph{Knowledge-grounded runtime graph}: a representation of lineage, dependency, validation, composition, mutation, and governance relations among evolving artifacts.

		\item \emph{Governance-oriented runtime architecture}: an architectural interpretation in which specialized agents coordinate generation, validation, evaluation, governance, and execution.
	\end{itemize}

	The paper therefore reframes runtime evolution not as unconstrained self-modification, but as a bounded, observable, and governable optimization process operating under explicit operational constraints.

	\section{Background and Related Work}

	Recent advances in agentic systems increasingly blur the distinction between generated code, runtime infrastructure, and persistent operational behavior. Emerging directions such as code-based agents, prompt and workflow optimization, executable harnesses, and runtime adaptation progressively treat generated artifacts not merely as transient outputs, but as reusable operational entities participating directly in future execution loops. This section situates the proposed framework within these converging research directions and highlights the gap between capability accumulation and governed runtime evolution.

	\subsection{Code as Agent Harness}

	The Code as Agent Harness perspective provides a broad taxonomy for understanding how code enters the agent loop. It organizes the literature into three connected layers: harness interface, harness mechanisms, and scaling the harness \citep{ning2026codeagentharness}. At the interface layer, code connects agents to reasoning, action, and environment modeling. At the mechanism layer, code supports planning, memory, tool use, control, and optimization for long-horizon execution. At the scaling layer, shared code artifacts, execution states, repositories, tests, and structured workflows support multi-agent coordination, review, and verification.

	Our proposal builds directly on that view but changes the center of gravity. The reference perspective primarily characterizes how code functions within agent harnesses. This paper focuses on how agent-initiated code artifacts can evolve as governed operational capabilities. In other words, we move from describing code as the medium of harnessed agent behavior to modeling the optimization dynamics of the artifacts that agents initiate inside the harness.

	This distinction matters because an agent-initiated artifact may play different roles over time. A generated test can first serve as a local verifier, then become a regression artifact, and later participate in future benchmark selection. A temporary tool can support one task and later become a reusable skill. A workflow can begin as a one-off execution plan and later become a persistent orchestration template. Once this occurs, the artifact is no longer merely a local aid for task completion. It has become part of the runtime's operational cognition.

	\subsection{Code-Centric Reasoning, Acting, and Environment Modeling}

	Several lines of work already show that executable artifacts can improve reasoning, action, and environment modeling. Program-aided reasoning methods externalize computation into executable programs \citep{chen2022program,gao2023pal,li2023chainofcode}, while ReAct-style agents interleave reasoning and action through tool-mediated observations \citep{yao2023react}. In embodied and interactive settings, generated programs can become action policies or reusable skill interfaces \citep{ahn2022saycan,liang2023codepolicies,wang2023voyager}. Code-grounded environments and benchmarks, including SWE-bench and agent software-engineering platforms, evaluate agents through executable state transitions, tests, and repository-level feedback \citep{jimenez2024swebench,openhands}.

	Voyager is particularly relevant because it treats executable code as an ever-growing skill library for lifelong learning. The agent generates executable programs, refines them with environment feedback and execution errors, stores successful skills, retrieves them in future tasks, and composes more complex behaviors from previously acquired ones \citep{wang2023voyager}. In this paper, Voyager is not used as evidence for governed runtime evolution, but as an important precursor showing that executable code can act as reusable operational memory.

	Recent work on skill optimization further strengthens this interpretation. SkillOpt treats an agent skill as an external trainable state for a frozen agent: trajectories are converted into bounded textual edits, candidate updates are accepted only through held-out validation, rejected edits are retained as negative feedback, and the final output is a compact reusable \texttt{best\_skill.md} artifact \citep{yang2026skillopt}. This provides empirical evidence that procedural artifacts can be optimized, exported, and transferred across models and execution harnesses. In the present paper, such optimized skills are interpreted as one important class of persistent operational artifact, while our focus remains broader: how these and other generated artifacts are governed, related, mutated, audited, and integrated into future runtime behavior.

	\subsection{Harness Engineering and Runtime Infrastructure}

	Modern agent frameworks such as LangGraph and DeepAgents provide durable execution, tool orchestration, persistent state, subagent delegation, and filesystem-like operational substrates \citep{langgraph,deepagents}. These frameworks operationalize many mechanisms described in Code as Agent Harness: they connect model outputs to tools, state, sandboxes, memories, and feedback loops. However, they do not by themselves define a theory of how agent-initiated artifacts should evolve, compete, persist, mutate, or become trusted operational capabilities.

	This paper treats these frameworks as infrastructure rather than as the main contribution. The proposed contribution is an architectural layer over such infrastructure: a way to model agent-generated artifacts as governed runtime capabilities whose lifecycle can be evaluated, traced, and optimized.

	\subsection{Self-Adaptive Systems and Autonomic Computing}

	The proposed architecture also relates to self-adaptive systems and autonomic computing.
	The MAPE-K loop established a classical model for systems that monitor, analyze, plan, execute, and maintain knowledge about themselves \citep{kephart2003autonomic,cheng2009selfadaptive}. The present work shares this concern with adaptation and operational knowledge, but differs in the nature of the adaptive substrate. In agentic systems, the substrate is not only configuration or policy; it also includes executable artifacts produced by the agents themselves during reasoning and action.

	\section{The Artifact Evolution Problem}

	The Code as Agent Harness view identifies agent-initiated code artifacts as underexplored, interactive code objects inside task loops \citep{ning2026codeagentharness}. We argue that this underexplored region defines a specific systems problem: the \emph{artifact evolution problem}.

	An agent-initiated artifact can be useful locally while still being unsafe, brittle, redundant, or misleading if persisted globally. Conversely, a temporary artifact may encode a valuable operational pattern that should be preserved and reused. The runtime therefore needs criteria and mechanisms for deciding whether an artifact should remain local, become part of memory, enter a candidate pool, be evaluated as a capability, be promoted into trusted reuse, or be deprecated.

	This problem cannot be solved by memory alone. Vector retrieval, summaries, and conversational histories preserve information, but they do not directly preserve executable behavior. It also cannot be solved by tool calling alone, because tool calling assumes a relatively stable set of developer-defined capabilities. Agent-initiated artifacts occupy the space between memory and tools: they are generated during execution, but may become future capabilities.

	This paper therefore frames artifact evolution as a harness-level optimization problem. The runtime must evaluate not only whether a candidate solves the current task, but whether it produces reusable executable cognition that improves future behavior without compromising safety, reproducibility, or governance.

	\section{From Agent-Initiated Artifacts to Executable Operational Cognition}

	We define \emph{executable operational cognition} as the persistent operational representation of executable artifacts that can influence future agent behavior, evaluation policies, runtime orchestration, and coordination.

	This definition deliberately extends the notion of agent-initiated code artifacts. The reference work emphasizes that agents create, execute, observe, revise, persist, and share code objects inside task loops \citep{ning2026codeagentharness}. Our proposal adds that once these objects persist and shape future decisions, they should be treated as cognitive runtime components. They are no longer only outputs, nor only tools. They become operational memory with executable semantics.

\begin{figure}[t]
\centering
\resizebox{0.96\textwidth}{!}{%
\begin{tikzpicture}[
	node distance=0.78cm and 1.0cm,
	every node/.style={font=\sffamily}
]

\node (task)
[layer_box, fill=white, minimum width=3.05cm, minimum height=0.82cm]
{\textbf{\faTasks\ Task Loop}\\[-1pt]{\scriptsize Local execution context}};

\node (artifact)
[layer_box, fill=layerArtifact, minimum width=3.05cm, minimum height=0.82cm, right=of task]
{\textbf{\faFileCode\ Agent Artifact}\\[-1pt]{\scriptsize Code, tests, tools, workflows}};

\node (execute)
[layer_box, fill=layerKernel, minimum width=3.05cm, minimum height=0.82cm, right=of artifact]
{\textbf{\faPlayCircle\ Execute--Revise}\\[-1pt]{\scriptsize Feedback-driven refinement}};

\node (persist)
[layer_box, fill=layerArtifact, minimum width=3.05cm, minimum height=0.82cm, right=of execute]
{\textbf{\faSave\ Persist--Evaluate}\\[-1pt]{\scriptsize Candidate capability}};

\node (cognition)
[core_box, fill=layerGov, minimum width=4.35cm, minimum height=0.98cm, below=1.0cm of $(artifact)!0.5!(execute)$]
{\textbf{\faBrain\ Executable Operational Cognition}\\[-1pt]{\scriptsize Persistent capability shaping future behavior}};

\node (mutation)
[layer_box, fill=layerGov, minimum width=2.85cm, minimum height=0.76cm, below=0.86cm of cognition, xshift=-3.85cm]
{\textbf{\faCodeBranch\ HarnessMutation}\\[-1pt]{\scriptsize Governed change}};

\node (lifecycle)
[layer_box, fill=layerKernel, minimum width=2.85cm, minimum height=0.76cm, below=0.86cm of cognition]
{\textbf{\faRedo\ Lifecycle}\\[-1pt]{\scriptsize Validate, promote, deprecate}};

\node (graph)
[layer_box, fill=layerBase, minimum width=2.85cm, minimum height=0.76cm, below=0.86cm of cognition, xshift=3.85cm]
{\textbf{\faProjectDiagram\ Runtime Graph}\\[-1pt]{\scriptsize Lineage and relations}};

\node (future)
[core_box, fill=layerBase, minimum width=4.15cm, minimum height=0.88cm, below=0.95cm of lifecycle]
{\textbf{\faCubes\ Future Runtime Behavior}\\[-1pt]{\scriptsize Governed capability reuse}};

\draw[gov_arrow] (task) -- (artifact);
\draw[gov_arrow] (artifact) -- (execute);
\draw[gov_arrow] (execute) -- (persist);
\draw[gov_arrow] (persist.south) |- (cognition.east);

\draw[gov_arrow] (cognition.south) -- ++(0,-0.28) -| (mutation.north);
\draw[gov_arrow] (cognition) -- (lifecycle);
\draw[gov_arrow] (cognition.south) -- ++(0,-0.28) -| (graph.north);

\draw[gov_arrow] (mutation.south) |- (future.west);
\draw[gov_arrow] (lifecycle) -- (future);
\draw[gov_arrow] (graph.south) |- (future.east);

\draw[trace_arrow]
([xshift=-0.18cm]future.west)
-- ++(-4.15cm,0)
|- ([yshift=-0.34cm]task.south)
node[pos=0.26,left,font=\scriptsize\bfseries] {Operational feedback};

\node[font=\scriptsize, align=center, below=0.16cm of future] {
	Governed persistence converts local artifacts into reusable runtime capabilities.
};

\end{tikzpicture}%
}
\caption{From agent-initiated artifacts to executable operational cognition. Local artifacts become persistent runtime capabilities only after evaluation, governance, lifecycle management, graph grounding, and reuse.}
\label{fig:artifact_lifecycle}
\end{figure}
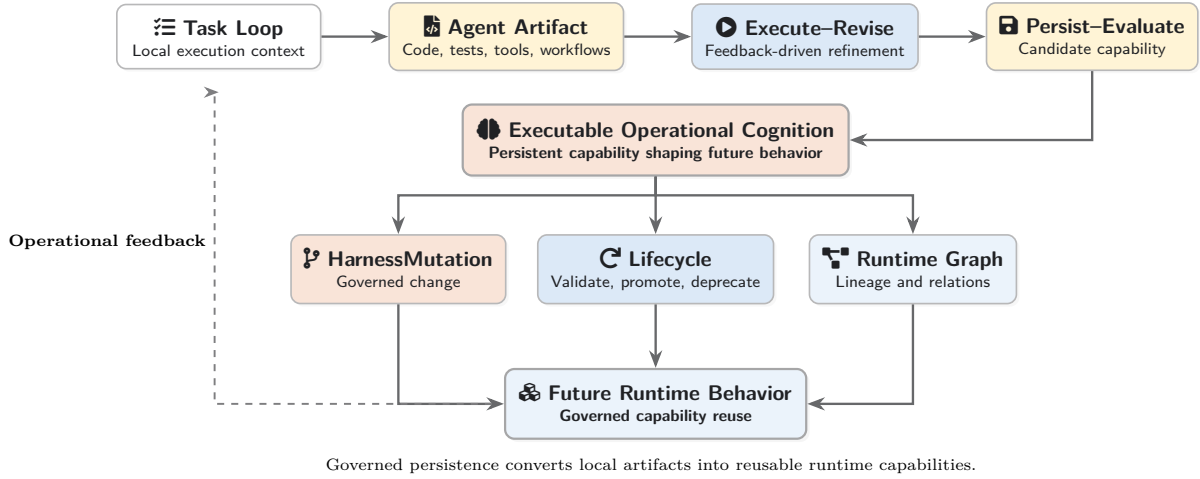

	This transition can be understood across three levels. At the first level, code is produced as a local task artifact, such as a script, test, or temporary utility. At the second level, the artifact participates in feedback-driven execution, enabling the agent to inspect behavior, revise actions, and verify progress. At the third level, the artifact is preserved, evaluated, governed, related to other artifacts, and reused as part of the future operational substrate. Figure~\ref{fig:artifact_lifecycle} summarizes this transition from local artifact production to governed runtime capability reuse. The third level is the focus of this paper.

	This framing also clarifies the role of memory. Many agent systems treat memory as retrieval over text, embeddings, summaries, or traces. In contrast, executable operational cognition treats memory as an active capability substrate where the runtime preserves executable structures that can be invoked, tested, revised, composed, and audited.

	\section{Harness-Oriented Operational Model}

	We define an executable harness as a governed operational unit:

	\begin{equation}
		H = \{P, T, E, M, G, O, K\},
	\end{equation}

	where $P$ denotes prompting policies, $T$ executable tools, $E$ evaluators, $M$ memory and contextual state, $G$ governance constraints, $O$ executable operational artifacts, and $K$ structured operational knowledge. The last two components are essential: $O$ represents agent-initiated artifacts that may persist beyond the task in which they were generated, while $K$ represents the knowledge-grounded structure that relates such artifacts to evaluations, dependencies, mutations, and runtime contexts.

	A concrete harness instance is represented as:

	\begin{equation}
		h_i = (p_i, t_i, e_i, m_i, g_i, o_i, k_i).
	\end{equation}

	Given a candidate set $\mathcal{C}_t = \{h_1, \ldots, h_k\}$ at iteration $t$, the runtime evaluates competing operational candidates under a multi-dimensional governance-aware objective.

	\begin{equation}
		h^* = \arg\max_{h_i \in \mathcal{C}_t} F(h_i),
	\end{equation}

	subject to operational constraints associated with cost, safety, robustness, reproducibility, and governance. The objective $F$ may combine task quality, validation strength, operational robustness, cost efficiency, and capability reuse:

	\begin{equation}
		F(h_i) = \alpha Q(h_i) + \beta R(h_i) + \gamma V(h_i) + \delta U(h_i) - \lambda C(h_i),
	\end{equation}

	where $Q$ denotes task quality, $R$ robustness, $V$ validation consistency, $U$ reuse value of generated operational artifacts, and $C$ operational cost.

	The addition of $U(h_i)$ distinguishes this formulation from conventional task-level optimization. A harness configuration may be valuable not only because it solves the current task, but also because it generates reusable executable cognition artifacts that improve future runtime behavior. Table~\ref{tab:positioning} summarizes how this shift positions the proposed model relative to the Code as Agent Harness view.

	\begin{table}[t]
		\centering
		\caption{Positioning of this paper relative to the Code as Agent Harness view.}
		\label{tab:positioning}
		\begin{tabular}{p{0.30\linewidth}p{0.30\linewidth}p{0.30\linewidth}}
			\toprule
			Dimension & Code as Agent Harness view & This paper \\
			\midrule
			Primary focus & Code as executable, inspectable, and stateful harness medium & Evolution and governance of persistent executable cognition artifacts \\
			Core object & Agent-initiated code artifacts inside the task loop & Harness configurations and lifecycle-managed operational capabilities \\
			Key mechanism & Generate--execute--observe--revise loops & HarnessMutation, selection, lifecycle promotion, runtime graph, observability \\
			Multi-agent role & Shared code artifacts for coordination and verification & Shared cognitive operational substrate with governed artifact evolution \\
			Open problem addressed & Reliable closed-loop code-harness behavior & Regression-aware optimization of artifact evolution over time \\
			\bottomrule
		\end{tabular}
	\end{table}

	\section{Governed Artifact Evolution}

	The proposed framework treats agent-generated artifacts as persistent operational entities participating directly in future runtime behavior. Unlike traditional execution pipelines where generated outputs are transient and disposable, evolving agent runtimes increasingly accumulate prompts, workflows, evaluators, routing rules, executable skills, and policies as reusable operational capabilities. This shift introduces a new systems problem: runtime evolution can no longer be understood as isolated execution, but as a governed lifecycle process involving validation, persistence, mutation, promotion, and operational oversight.

	Rather than framing adaptation as unrestricted self-modification, the proposed model treats runtime evolution as a bounded and observable process operating under explicit governance constraints. The following subsections introduce the lifecycle model, governed mutation mechanisms, and capability-selection structures enabling controlled evolution of persistent operational artifacts.

	\subsection{Harness Mutation}

	To model controlled evolution of the operational substrate, we introduce \emph{HarnessMutation}. A HarnessMutation represents a governed transformation over harness configurations and executable operational artifacts:

	\begin{equation}
		\mu : h_i \rightarrow h'_i.
	\end{equation}

	A mutation may affect prompts, evaluator policies, orchestration workflows, routing strategies, retrieval behavior, memory compaction rules, reusable skills, benchmark definitions, or runtime graph relations. The critical point is that these transformations are not unrestricted self-modifications. They are explicit operational changes applied to the runtime substrate and should therefore be observable, versioned, validated, reversible, and governance-aware.

	This perspective extends the harness-evolution problem identified in prior work, where automatic runtime adaptation can overfit, weaken safety guarantees, increase operational cost, hide regressions, or silently degrade reliability under distributional shifts \citep{ning2026codeagentharness}. Treating HarnessMutation as a first-class runtime object makes these risks structurally explicit.

	Skill-level optimization offers a concrete example of this kind of bounded mutation. In SkillOpt, edits to an external skill document are constrained by textual edit budgets, evaluated through held-out gates, and rejected when they fail to improve validation performance \citep{yang2026skillopt}. From the perspective of this paper, such skill edits can be viewed as a specialized form of HarnessMutation: they modify one component of the operational substrate under explicit update, validation, and rejection rules. The broader runtime problem is to generalize this discipline beyond skills to evaluators, workflows, policies, routing behavior, memory rules, graph relations, and capability lifecycle state.

	Under this interpretation, mutations are not directly adopted into future cognition. Instead, they remain governed candidates subject to lifecycle-aware evaluation and operational review. Each mutation should therefore carry a bounded change contract specifying: the operational component modified, the targeted failure mode, the expected improvement, the invariants preserved, the evaluation capable of falsifying the change, and the rollback conditions required for safe recovery.

	\subsection{Capability Lifecycle}

	Generated operational capabilities evolve through explicit lifecycle states:

	\begin{equation}
		L = \{\text{experimental}, \text{validated}, \text{trusted}, \text{canonical}, \text{deprecated}\}.
	\end{equation}

	This lifecycle model separates capability generation from capability adoption. An artifact may initially emerge from local task execution, later undergo validation through execution traces, evaluators, or governance review, and eventually become part of the persistent operational substrate reused by future runtime executions.

	The lifecycle structure also introduces bounded operational stability. Capabilities can be promoted into trusted reuse, stabilized as canonical operational cognition, or deprecated when evidence suggests drift, redundancy, unsafe behavior, excessive operational cost, or declining utility. Runtime evolution therefore becomes an evidence-driven governance process rather than unrestricted accumulation of generated artifacts.

	\subsection{Lifecycle-Governed Capability Selection}

	A common limitation of single-trajectory refinement is that the runtime commits early to one operational path. Search-based planning and tree-style exploration have already demonstrated the value of comparing alternative execution trajectories \citep{li2025codetree}, while multi-agent systems such as AgentCoder and MapCoder distribute planning, generation, testing, and debugging across specialized roles \citep{huang2023agentcoder,islam2024mapcoder}. The proposed framework extends this perspective toward lifecycle-governed selection over persistent runtime capabilities and harness mutations.

	Given a candidate set $P_t = \{h_1, \ldots, h_k\}$ at iteration $t$, the runtime evaluates competing operational candidates under a multi-dimensional governance-aware objective. Rather than relying exclusively on immediate task success, evaluation should incorporate execution reliability, verifier quality, reproducibility, operational cost, safety constraints, graph consistency, composability, and long-term reuse potential.

	This perspective is particularly important for persistent agent-generated artifacts. Two workflow variants, evaluators, reusable skills, or routing policies may both satisfy the immediate task objective while differing substantially in robustness, maintainability, operational risk, or future adaptability. Lifecycle-governed selection therefore treats generated artifacts not merely as local outputs, but as competing operational capabilities whose promotion influences future runtime cognition and behavior.

	Under this interpretation, runtime evolution becomes neither unrestricted self-modification nor static configuration management. Instead, it becomes a governed process of bounded operational selection, where only validated and lifecycle-consistent capabilities progressively shape the persistent cognitive substrate.

	\section{Knowledge-Grounded Operational Cognition}

	Once artifacts persist and evolve, the runtime requires a structured way to represent how they relate to one another. A generated validator may depend on a workflow, a benchmark may validate a skill, a mutation may supersede an older artifact, and a policy may constrain when a capability can be invoked. Without an explicit representation of these relations, persistent artifacts risk becoming an unstructured capability library.

	We therefore introduce a \emph{Knowledge-Grounded Runtime Graph} as a structured operational memory layer. The graph is not intended to replace vector retrieval or trace storage. Its role is to represent the operational and epistemic relations that make artifact evolution governable.

\begin{figure}[t]
\centering
\begin{tikzpicture}[node distance=0.88cm]

\node (agents)
[layer_box, fill=layerArtifact, minimum width=8.25cm, minimum height=1.08cm]
{\begin{tabular}{c}
\textbf{\faUsers\ Specialized Governance Agents}\\[1pt]
{\scriptsize Generation, validation, review, reflection}
\end{tabular}};

\node (kernel)
[core_box, fill=layerGov, minimum width=8.25cm, minimum height=1.08cm, below=of agents]
{\begin{tabular}{c}
\textbf{\faUserShield\ Governed Runtime Kernel}\\[1pt]
{\scriptsize Lifecycle control, rollback, audit trail, risk gates}
\end{tabular}};

\node (graph)
[layer_box, fill=layerKernel, minimum width=8.25cm, minimum height=1.08cm, below=of kernel]
{\begin{tabular}{c}
\textbf{\faProjectDiagram\ Knowledge-Grounded Runtime Graph}\\[1pt]
{\scriptsize Lineage, dependencies, validation, mutation history, capability relations}
\end{tabular}};

\node (substrate)
[layer_box, fill=layerBase, minimum width=8.25cm, minimum height=1.08cm, below=of graph]
{\begin{tabular}{c}
\textbf{\faLayerGroup\ Execution and Artifact Substrate}\\[1pt]
{\scriptsize Tools, state, sandboxes, traces, policies, persistent artifacts}
\end{tabular}};

\node (obs)
[layer_box, fill=white, minimum width=2.52cm, minimum height=1.0cm, anchor=west]
at ([xshift=1.05cm]graph.east)
{\begin{tabular}{c}
\textbf{\faEye\ Observability}\\[-1pt]
{\scriptsize Telemetry}\\[-1pt]
{\scriptsize Lineage traces}
\end{tabular}};

\draw[gov_arrow] (agents) -- (kernel);
\draw[gov_arrow] (kernel) -- (graph);
\draw[gov_arrow] (graph) -- (substrate);

\draw[trace_arrow] (substrate.east) -| (obs.south);
\draw[trace_arrow] (obs.north) |- (agents.east);
\draw[trace_arrow] (obs.west) -- (graph.east);

\end{tikzpicture}
\caption{Conceptual knowledge-grounded runtime architecture. Governance-aware layers coordinate specialized agents, lifecycle control, runtime-graph memory, execution artifacts, and observability without committing to a specific implementation framework.}
\label{fig:kg_architecture}
\end{figure}
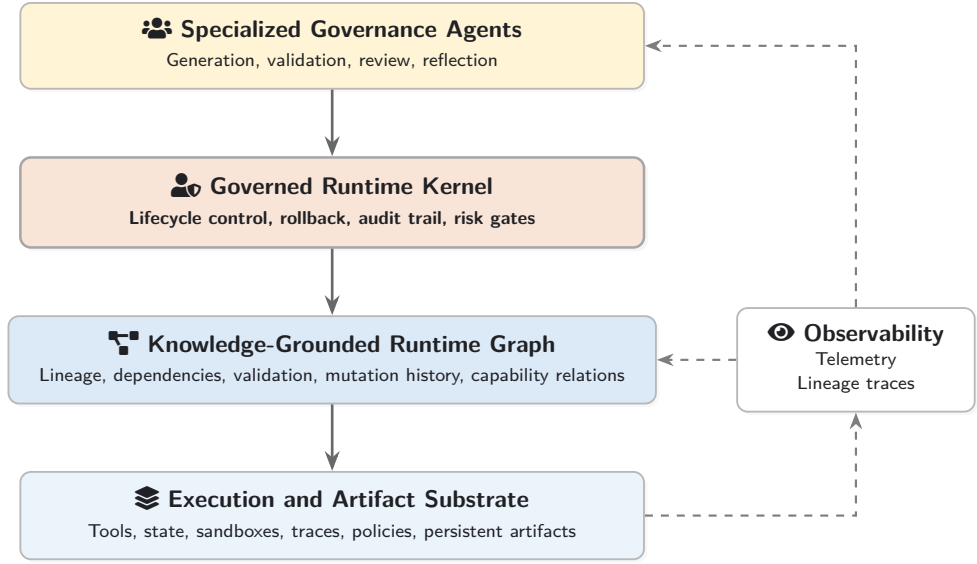

	Figure~\ref{fig:kg_architecture} introduces the graph at the conceptual architectural level. Its purpose is structural: it shows where the graph sits between specialized governance agents, the governed runtime kernel, and a generic execution-and-artifact substrate. The figure deliberately avoids naming a concrete framework; implementation-specific runtimes such as LangGraph and DeepAgents are introduced later in Figure~\ref{fig:deepagents_reference_architecture}. Here, the goal is to clarify the persistent memory layer that makes lineage, dependency, validation, mutation history, and observability available to governance mechanisms.

	\FloatBarrier

	Let the runtime graph at time $t$ be:

	\begin{equation}
		\mathcal{G}_t = (V_t, E_t),
	\end{equation}

	where $V_t$ denotes operational entities and $E_t$ typed relations. Nodes may represent agents, skills, workflows, evaluators, policies, benchmarks, datasets, traces, and mutations. Edges encode relations such as dependency, provenance, validation, composition, improvement, supersession, failure, and mutation lineage:

	\begin{equation}
		E_t \subseteq V_t \times R \times V_t,
	\end{equation}

	with

	\begin{equation}
		\begin{aligned}
			R = \{&\texttt{depends\_on},\texttt{generated\_by},\texttt{validated\_by},
			\texttt{improves},\texttt{supersedes},\texttt{mutated\_from},\\
			&\texttt{composed\_with},\texttt{fails\_under}\}.
		\end{aligned}
	\end{equation}

	Each operational entity $v_i \in V_t$ may be represented by:

	\begin{equation}
		\phi(v_i) = (c_i, q_i, \tau_i, \ell_i),
	\end{equation}

	where $c_i$ denotes executable content or specification, $q_i$ an operational quality score, $\tau_i$ temporal and lifecycle metadata, and $\ell_i$ lineage information. The quality score can integrate performance, robustness, stability, reuse utility, and governance risk. A simple scalarized form is:

	\begin{equation}
		q_i = \omega_p p_i + \omega_r r_i + \omega_s s_i + \omega_u u_i - \omega_\rho \rho_i,
	\end{equation}

	where $p_i$ measures performance, $r_i$ robustness, $s_i$ stability, $u_i$ reuse utility, and $\rho_i$ operational risk. The coefficients allow the runtime to express different operational priorities.

	The graph enables graph-grounded capability composition. Given a set of available skills $\mathcal{S}_t = \{s_1,\ldots,s_n\}$, the runtime may synthesize a composed capability:

	\begin{equation}
		\Psi : \mathcal{P}(\mathcal{S}_t) \rightarrow s^*,
	\end{equation}

	where composition decisions are guided by dependency relations, validation histories, benchmark coverage, and previous mutation outcomes. This makes composition less ad hoc: the runtime can reuse operational knowledge about which artifacts work together, under which contexts, and with which failure modes.

	The graph also introduces an epistemic governance layer. Persisting an artifact therefore means preserving explicit claims about validity, scope, provenance, and expected behavior. The Knowledge-Grounded Runtime Graph makes such claims inspectable, contestable, and governable.

	Figure~\ref{fig:runtime-governance-loop} then complements the structural view with a temporal view. It focuses on the governed evolution loop: artifacts are generated, evaluated against traces and benchmarks, reviewed under risk and approval constraints, staged as HarnessMutation proposals, and only then promoted into persistent runtime behavior. This separates the role of Figure~\ref{fig:kg_architecture} from Figure~\ref{fig:runtime-governance-loop}: the former explains the architectural placement of the graph, whereas the latter explains how capabilities move through governance over time.

	\begin{figure}[t]
\centering
\begin{tikzpicture}[
	node distance=0.72cm and 1.25cm,
	every node/.style={font=\sffamily}
]

\node (gen)
[layer_box, fill=layerArtifact, minimum width=4.7cm, minimum height=0.78cm]
{\textbf{\faLightbulb\ Generate Artifact}\\[-1pt]{\scriptsize Skill, workflow, evaluator, policy}};

\node (eval)
[layer_box, fill=layerKernel, minimum width=4.7cm, minimum height=0.78cm, below=of gen]
{\textbf{\faCheckDouble\ Evaluate Candidate}\\[-1pt]{\scriptsize Traces, tests, benchmarks, alternatives}};

\node (review)
[core_box, fill=layerGov, minimum width=4.7cm, minimum height=0.78cm, below=of eval]
{\textbf{\faBalanceScale\ Governance Review}\\[-1pt]{\scriptsize Evidence, risk gates, approval}};

\node (mutation)
[layer_box, fill=layerGov, minimum width=4.7cm, minimum height=0.78cm, below=of review]
{\textbf{\faCodeBranch\ Stage HarnessMutation}\\[-1pt]{\scriptsize Change contract and rollback conditions}};

\node (promote)
[core_box, fill=layerBase, minimum width=4.7cm, minimum height=0.86cm, below=of mutation]
{\textbf{\faRocket\ Promote or Deprecate}\\[-1pt]{\scriptsize Persistent operational capability}};

\node (graph)
[core_box, fill=layerKernel, minimum width=3.7cm, minimum height=2.05cm, right=1.05cm of review]
{\textbf{\faProjectDiagram\ Runtime Graph}\\[-1pt]
{\scriptsize Lineage}\\[-1pt]
{\scriptsize Validation evidence}\\[-1pt]
{\scriptsize Mutation history}};

\draw[gov_arrow] (gen) -- node[midway, right, font=\scriptsize\bfseries] {execute} (eval);
\draw[gov_arrow] (eval) -- node[midway, right, font=\scriptsize\bfseries] {select} (review);
\draw[gov_arrow] (review) -- node[midway, right, font=\scriptsize\bfseries] {approve} (mutation);
\draw[gov_arrow] (mutation) -- node[midway, right, font=\scriptsize\bfseries] {integrate} (promote);

\draw[trace_arrow] (eval.east) -- (graph.west);
\draw[trace_arrow] (review.east) -- (graph.west);
\draw[trace_arrow] (mutation.east) -- (graph.west);
\draw[trace_arrow] (graph.south) |- (promote.east);

\draw[trace_arrow]
(promote.west)
to[out=180,in=180,looseness=1.18]
node[pos=0.54,left,font=\scriptsize\bfseries] {reuse feedback}
(gen.west);

\node[font=\scriptsize, align=center, text=textDark!76, below=0.18cm of promote] {
	Temporal view: only reviewed changes become reusable operational behavior.
};

\end{tikzpicture}
\caption{Governed runtime evolution loop. Agent-generated artifacts move through evaluation, governance review, mutation staging, graph recording, and promotion or deprecation before influencing future runtime behavior.}
\label{fig:runtime-governance-loop}
\end{figure}

	\FloatBarrier

	\section{Prototype Architecture over Modern Agent Runtimes}

	The proposed architecture can be implemented over existing agent-runtime primitives rather than by building a complete framework from scratch. Modern runtimes such as LangGraph and DeepAgents already provide durable execution, tool orchestration, state persistence, middleware control, subagent delegation, filesystem-like operational state, and long-horizon execution primitives \citep{langgraph,deepagents}. These systems operationalize many mechanisms described in the Code as Agent Harness perspective because they expose the runtime structures through which models interact with tools, execution environments, memory, and feedback loops.

	The implementation examples discussed in this section rely on recent runtime-oriented frameworks and technical documentation, including LangGraph and DeepAgents. They should therefore be interpreted as evolving operational infrastructures rather than as formally evaluated research systems. Here, they serve as practical execution substrates for illustrating how governance-oriented runtime layers can be integrated over modern agent infrastructures.

	Our proposal treats these runtimes as the underlying execution substrate while introducing an additional governance-oriented layer responsible for lifecycle management, mutation control, graph-grounded operational memory, observability, and capability promotion. Under this interpretation, generated skills, evaluators, workflows, policies, benchmarks, and mutation proposals are no longer transient execution outputs; they become persistent operational entities subject to validation, review, promotion, rollback, and runtime governance.
	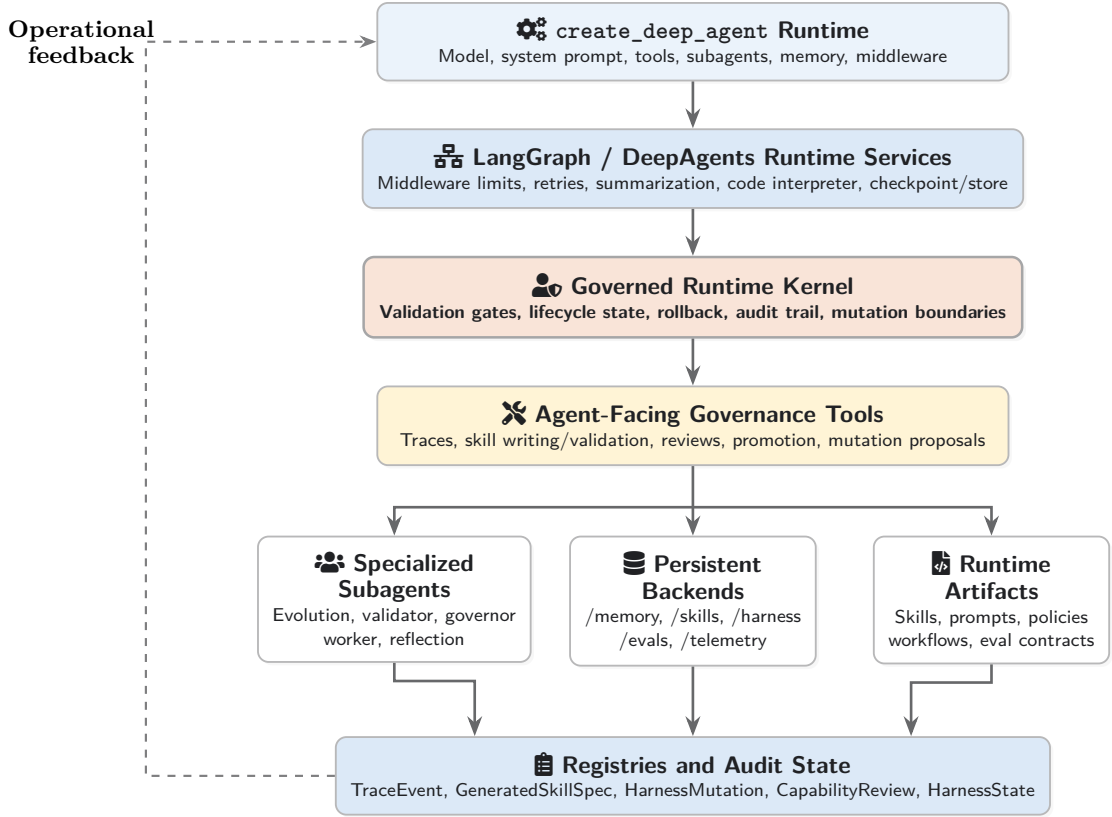
\begin{figure}[t]
\centering
\begin{tikzpicture}[node distance=0.62cm and 0.95cm]

\node (agent)
[layer_box, fill=layerBase, minimum width=8.35cm, minimum height=0.88cm]
{\textbf{\faCogs\ \texttt{create\_deep\_agent} Runtime}\\[-1pt]{\scriptsize Model, system prompt, tools, subagents, memory, middleware}};

\node (middleware)
[layer_box, fill=layerKernel, minimum width=8.35cm, minimum height=0.88cm, below=of agent]
{\textbf{\faNetworkWired\ LangGraph / DeepAgents Runtime Services}\\[-1pt]{\scriptsize Middleware limits, retries, summarization, code interpreter, checkpoint/store}};

\node (kernel)
[core_box, fill=layerGov, minimum width=8.35cm, minimum height=0.98cm, below=of middleware]
{\textbf{\faUserShield\ Governed Runtime Kernel}\\[-1pt]{\scriptsize Validation gates, lifecycle state, rollback, audit trail, mutation boundaries}};

\node (tools)
[layer_box, fill=layerArtifact, minimum width=8.35cm, minimum height=0.88cm, below=of kernel]
{\textbf{\faTools\ Agent-Facing Governance Tools}\\[-1pt]{\scriptsize Traces, skill writing/validation, reviews, promotion, mutation proposals}};

\node (fanout) [below=0.42cm of tools] {};

\node (subagents)
[layer_box, fill=white, minimum width=2.34cm, minimum height=0.96cm, below=0.92cm of tools, xshift=-3.95cm]
{\textbf{\faUsers\ Specialized}\\\textbf{Subagents}\\[-1pt]{\scriptsize Evolution, validator, governor}\\[-1pt]{\scriptsize worker, reflection}};

\node (backends)
[layer_box, fill=white, minimum width=2.34cm, minimum height=0.96cm, below=0.92cm of tools]
{\textbf{\faDatabase\ Persistent}\\\textbf{Backends}\\[-1pt]{\scriptsize /memory, /skills, /harness}\\[-1pt]{\scriptsize /evals, /telemetry}};

\node (artifacts)
[layer_box, fill=white, minimum width=2.34cm, minimum height=0.96cm, below=0.92cm of tools, xshift=3.95cm]
{\textbf{\faFileCode\ Runtime}\\\textbf{Artifacts}\\[-1pt]{\scriptsize Skills, prompts, policies}\\[-1pt]{\scriptsize workflows, eval contracts}};

\node (registries)
[layer_box, fill=layerKernel, minimum width=8.35cm, minimum height=1.02cm, below=0.92cm of backends]
{\textbf{\faClipboardList\ Registries and Audit State}\\[-1pt]{\scriptsize TraceEvent, GeneratedSkillSpec, HarnessMutation, CapabilityReview, HarnessState}};

\draw[gov_arrow] (agent) -- (middleware);
\draw[gov_arrow] (middleware) -- (kernel);
\draw[gov_arrow] (kernel) -- (tools);

\draw[draw=textDark!68, line width=1pt] (tools.south) -- (fanout.center);
\draw[gov_arrow] (fanout.center) -| (subagents.north);
\draw[gov_arrow] (fanout.center) -- (backends.north);
\draw[gov_arrow] (fanout.center) -| (artifacts.north);

\draw[gov_arrow] (subagents.south) -- ++(0,-0.24) -| ([xshift=1.85cm]registries.north west);
\draw[gov_arrow] (backends.south) -- (registries.north);
\draw[gov_arrow] (artifacts.south) -- ++(0,-0.24) -| ([xshift=-1.85cm]registries.north east);

\draw[trace_arrow]
(registries.west)
-- ++(-2.5cm,0)
|- (agent.west)
node[
	pos=0.54,
	left,
	align=center,
	font=\small\bfseries
]
{Operational\\feedback};

\end{tikzpicture}
\caption{Prototype architecture over modern agent runtimes. The governed runtime kernel operates over LangChain/DeepAgents infrastructure to coordinate lifecycle-aware persistence, mutation governance, validation, observability, and operational cognition management.}
\label{fig:deepagents_reference_architecture}
\end{figure}

	Figure~\ref{fig:deepagents_reference_architecture} summarizes the reference implementation structure in \texttt{governed\_agent\_runtime\_v10.py}. The architecture separates runtime orchestration from governance-aware operational persistence. LangGraph and DeepAgents provide the execution substrate, middleware coordination, subagent orchestration, and persistent runtime services, while the governed runtime kernel introduces explicit lifecycle boundaries over generated artifacts.

	Specialized agents participate in generation, validation, evaluation, governance review, reflection, and operational execution. Persistent registries maintain lifecycle state, mutation lineage, validation evidence, observability traces, and promotion history. In this way, the runtime preserves operational memory about how capabilities evolve over time rather than treating generated artifacts as isolated task outputs.

	This architecture should not be interpreted as a complete autonomous runtime framework. Rather, it is a governance-oriented operational layer over existing agent-runtime infrastructure. The proposal formalizes what happens after executable artifacts are generated: how they are validated, evaluated, persisted, promoted, related through operational graphs, mutated, audited, and eventually integrated into future runtime behavior.

	\section{Illustrative Operational Scenario}

	Consider a long-running software or research assistant repeatedly encountering schema inconsistencies in task inputs. During one task, an agent writes a temporary normalization script to transform inconsistent payloads into a validated structure. In a conventional workflow, this script may remain a local artifact and disappear after the task.

	Under the proposed architecture, the artifact enters a candidate lifecycle. It is executed, tested, traced, and compared against alternative normalizers. If it repeatedly improves validation consistency, it can be promoted into an experimental capability. Additional evaluations may then compare robustness under edge cases, cost, failure transparency, and compatibility with downstream tools. If the artifact remains useful, it becomes trusted or canonical; if it fails under drift, it is deprecated or mutated.

	The runtime graph records this evolution. The normalizer may be linked to the validator that approved it, the benchmark that tested it, the workflow that uses it, the mutation that improved it, and the failure cases that constrain its scope. The same logic applies to generated tests, evaluators, workflow templates, benchmark generators, and routing policies. The key idea is that operational experience is not only summarized; it is converted into executable and structured runtime capability.

	\section{Governance and Observability}

	Governance-aware observability is required because persistent executable artifacts can affect future behavior. If a generated evaluator is flawed, it can reward the wrong behavior. If a workflow mutation hides failures, it can improve apparent task success while reducing reliability. If a reusable skill is promoted too early, it can propagate brittle assumptions across future tasks.

	For this reason, the runtime should maintain operational lineage linking artifacts, evaluations, traces, graph relations, mutations, and lifecycle transitions. Observability should include not only final task outcomes but also prompts, tool calls, execution traces, evaluator results, sandbox state, cost, latency, human approvals, graph updates, and rollback events. This aligns with the reference work's emphasis on deep telemetry as a substrate for harness diagnosis and evolution \citep{ning2026codeagentharness}.

	Governance is therefore not an external policy layer added after the fact. It is part of the runtime semantics of executable operational cognition. To persist an artifact is to make a governance decision about future behavior; to add a graph relation is to make an epistemic claim about how that behavior should be interpreted.

	\section{Distributed Runtime Implications}

	The proposed perspective is especially relevant in distributed operational environments. As agentic systems scale across services, teams, repositories, data platforms, and orchestration layers, persistent artifacts become coordination structures as much as local execution units. A generated evaluator may be reused across services; a workflow may encode a cross-system operational policy; a mutation may affect shared runtime behavior; a graph relation may determine which capability is trusted under a specific context.

	This creates distributed systems challenges. Runtime cognition must remain consistent enough to be useful, but flexible enough to evolve. Capabilities may need local specialization while preserving shared provenance. Graph updates may require auditability, versioning, and rollback. Evaluators and benchmarks may drift as operational contexts change. These issues connect agentic runtime design with classical concerns in distributed computing: consistency, replication, fault isolation, observability, coordination, and cost-aware execution.

	This connection is important because it prevents the proposed framework from being reduced to a coding-agent abstraction. The deeper question is how agent-generated executable artifacts can become governed operational substrate in complex systems where adaptation, reuse, and coordination must be controlled over time.

	\section{Discussion}

	The proposed framework reframes agentic self-improvement as governed optimization over executable operational cognition. This perspective is narrower than unrestricted self-modification and broader than prompt optimization: it focuses on agent-initiated artifacts that may outlive a single task and later shape system behavior.

	The main conceptual shift is from code as task output to code as lifecycle-managed capability. The Code as Agent Harness view establishes the importance of executable, inspectable, and stateful code artifacts inside agent loops. This paper argues that the next systems problem is to manage those artifacts through mutation control, lifecycle-aware selection, observability, graph-grounded representation, and governed reuse.

	This framing also positions the proposal relative to adjacent work. Prompt evolution and context engineering optimize the information presented to the model \citep{zhang2025agenticcontext,agrawal2025gepa}. Harness optimization improves the external software layer around the model \citep{lee2026metaharness,lou2026autoharness}. Lifelong code-based agents and skill-optimization methods accumulate or refine reusable procedural capabilities \citep{wang2023voyager,lin2026uivoyager,yang2026skillopt}. The proposed framework targets the systems-level intersection of these directions by treating agent-initiated artifacts as operational capabilities whose evolution is explicitly governed, structurally represented, and integrated into runtime lifecycle control.

	Several open challenges remain. Evaluation signals may be incomplete or misleading, causing the runtime to promote artifacts that overfit weak benchmarks. Mutation operators may improve average performance while regressing rare cases. Multi-agent systems may share capabilities without consistent state convergence. Capability libraries may become redundant, stale, or unsafe. Runtime graphs may encode incorrect or outdated relations. These risks suggest that future work should focus on regression-aware mutation policies, artifact-level trust scores, graph consistency, benchmark adequacy, and safe exploration over harness configurations.

	\section{Limitations and Future Work}

	This paper is intentionally positioned as an architectural and conceptual contribution. It does not claim unrestricted recursive self-improvement, autonomous runtime self-redesign, or a fully validated AI Scientist system. It also does not provide a large-scale benchmark demonstrating that the proposed runtime graph and mutation mechanisms outperform existing systems across domains. These boundaries are important: the contribution is a governance model for artifact evolution, not an empirical claim that such evolution is already solved.

	These limitations are deliberate. The goal of the current paper is to formalize the artifact evolution problem and define a coherent substrate for future experimentation. Future work should evaluate concrete implementations across software engineering, research assistance, data workflows, and distributed operations. Particularly important directions include benchmark synthesis, regression-aware artifact promotion, graph-guided capability composition, mutation trust regions, distributed capability synchronization, and cost-aware runtime optimization.

	A further limitation is that the runtime graph is treated primarily as an architectural abstraction. Future implementations should examine how graph quality, lineage accuracy, evaluator reliability, and rollback constraints affect long-horizon robustness. This includes studying when graph-grounded reuse improves capability composition and when it propagates stale or unsafe operational assumptions.

	The broader implication is that future agentic infrastructures may increasingly resemble evolving operational ecosystems rather than static software pipelines. In such settings, executable artifacts become lifecycle-governed entities that influence future decisions, raising systems questions about optimization stability, distributed coordination, governance semantics, observability, and operational trust.

	\section{Conclusion}

	This paper proposed a harness-oriented framework for governing the evolution of agent-initiated code artifacts. Building on the Code as Agent Harness perspective, it focused on the underexplored transition from local generated artifacts to reusable capabilities that can shape later task execution.

	The central contribution is the notion of executable operational cognition: generated artifacts become governable capabilities when they are evaluated, mutated, promoted, deprecated, composed, related through a runtime graph, and reused under explicit lifecycle constraints. HarnessMutation operators, capability lifecycle management, lifecycle-governed selection, knowledge-grounded operational memory, and observability provide the main mechanisms for this controlled evolution.

	The broader implication is that future agent infrastructures may need to manage not only prompts, tools, and memory, but also the growing body of executable capabilities produced by agents themselves. Treating these capabilities as governed system components offers a path toward adaptive agent runtimes that remain auditable, reversible, and operationally constrained.

	Future agentic systems may therefore evolve not through model retraining alone, but through the governed accumulation, mutation, and orchestration of executable operational capabilities.

	\FloatBarrier

	\section*{Code Availability}

	The accompanying reference implementation operationalizes the proposed architecture as an executable reference substrate rather than as empirical validation. It instantiates the main conceptual objects introduced in the paper, including \texttt{TraceEvent}, \texttt{GeneratedSkillSpec}, \texttt{HarnessMutation}, \texttt{CapabilityReview}, governance policy constraints, lifecycle transitions, and agent-facing tools for trace inspection, mutation proposal, validation, review, and promotion.

	The reference implementation, governance specifications, evaluation pipelines, and observability integrations will be publicly released at:
	\begin{center}
		\url{https://github.com/mgarralda/governed-runtime}
	\end{center}

	\section*{Acknowledgements}
	The author acknowledges the Laboratorio de Innovación Aplicada (L2IA) at Minsait (Indra Group) for fostering exploration and research in AI systems, distributed runtimes, and applied agentic infrastructures.

	\bibliographystyle{unsrtnat}
	\bibliography{references}

\end{document}